\documentclass[12pt,preprint]{aastex}
\usepackage{lscape}

\begin{document}
\title{Wide Field Imager in Space for Dark Energy and Planets}

\author{Andrew Gould}

\affil{
Institut d'Astrophysique de Paris, Ohio State University\\
gould@astronomy.ohio-state.edu
}

\begin{abstract}

A wide-field imager in space could make remarkable progress in
two very different frontiers of astronomy: dark energy and extra-solar
planets.  Embedding such an imager on a much larger and more complicated
DE mission would be a poor science-approach under any circumstances
and is a prescription for disaster in the present fiscal climate.
The 2010 Decadal Committee must not lead the lemming stampede that
is driving toward a DE mega-mission, but should stand clearly in
its path.

\end{abstract}


\section{WMAP Model for DE: Faster, Cheaper, ``Better''
\label{FCB}}

Dark energy (DE) is arguably the most important physics problem
of the 21st century, with major implications for astronomy,
fundamental physics, and perhaps even philosophy.  Unfortunately,
a variety of bureaucratic and sociological forces on several
continents are now driving toward a dark energy mega-mission
that would simultaneously attack this problem on 3 fronts:
weak lensing (WL), baryon acoustic oscillations (BAO) and 
supernovae (SN).
If adopted, this course of action will produce an utter disaster,
delaying progress on a crucial frontier of science for many
decades.   While the science goals of these 3 experiments
are complementary,
the instrumentation is not, and hence the costs and engineering
complexity are bound to spiral out of control.  Moreover, we
are entering an era of severe financial crisis when such
exponentiating costs simply will not be tolerated.

The siren call leading to this disaster is that only by obtaining
agreement among 3 independent DE measurements, each with its own
systematics, will it be possible to solve the DE problem.
This is nonsense: DE will not be ``solved'' by
this mega-mission, nor 2 or 3 of them.  It will dominate 21st century
physics.  The missions currently conceived will at best offer some
initial clues.  

{\it WMAP} offers a far better model for attacking such a scientifically
compelling and technologically challenging problem: faster, cheaper,
``better''.  I have put ``better'' in quotes because while {\it WMAP}
was fast and cheap, {\it Planck} will obviously be better. But from
the standpoint of advancing {\it CMB} science in the broadest sense,
including practical development of the field, theoretical inquiry,
and -- very importantly -- motivation and design of future projects,
{\it WMAP}'s rapid launch and solid data actually did make it ``better''
than waiting a decade for a ``better'' satellite.

These lessons apply even more strongly to DE.  We are struggling to
measure 1 or 2 parameters, not refine a basically coherent model.
Hence, what we learn at each stage is even more crucial to the
design of the next.  There is no doubt that any results from
a single experiment will be called into question on account of
systematics, but that is not reason to delay simple experiments
in favor of more complicated, later ones.  On the contrary, the
doubt raised by early experiments will be the strongest driver
to construct new, more decisive experiments.

This was true for the SN results that first put DE on solid footing.
It was also true for the Bahcall-Davis solar-neutrino experiment.
They didn't wait for the massive, technically challenging, and very
expensive $p$-$p$ experiments to be feasible: they tested the
solar model with what was accessible with 1960s technology and funding.
Yes, their results were doubted for 30 years, but these doubts are
exactly what drove future experiments.  Science is about doubt.

The 2010 Decadal Committee faces a choice:  stand at the head
of the DE lemmings and recommend a mission that will never be built,
or stand in their path and recommend a mission that is faster, and
cheaper, and, yes, ``better''.

\section{DE and Planets: Convergent Evolution
\label{converge}}

The upshot of \S~\ref{FCB} is that the first DE satellite should
be simple and attack DE by one of the routes (WL, BAO, SN), not
all three.  No argument was made as to which one.  And from the
standpoint of DE alone, I do not think that a compelling choice
can be made among these three.  For example, on DE grounds alone,
a reasonably good case could be made for ADEPT, which is a 
relatively simple mission aimed primarily at BAO.

But here I want to point out a very simple fact:  The wide-field
imaging satellite needed to do a weak-lensing survey is essentially
identical to the one needed to do a microlensing planet search.

Here I do not mean that back-of-the-envelope calculations lead to
the same sort of figures of merit.  I mean literally that two
{\it completely independent} efforts were made to design
satellites that would achieve these very different goals, and
the characteristics derived from fairly mature engineering 
work were almost identical: same aperture, same IR pixel scale,
same IR camera size, same orbit.  The only significant difference
is that the WL satellite requires an optical camera in addition
to IR.  Even the fields are complementary:  WL looks at high-latitude
fields while microlensing looks at the Galactic bulge.  So they could
amicably share time on the same satellite, with microlensing
observations taking place during the 3 months per year that 
the camera cannot view high-latitude fields and so must observe
Galactic-plane targets.

When I say ``completely independent'', I mean that there was 
absolutely no contact between the science/engineering teams that
developed these designs.  Indeed they did not even know of each
others' existence until a French astronomer in contact with both
DUNE (WL) and MPF (microlensing) put them in contact with each other.

I will not go into detail about the designs.  This is properly the
subject of an RFI paper.  Here I am just focusing on the proper
scientific approach to two big problems.

\section{Microlensing Planet Searches
\label{ulensing}}

Microlensing is potentially the most powerful method of
finding planets.  It is the only method sensitive to analogs
of all $M>0.1\,M_\oplus$ solar-system planets (no method is
sensitive to Mercuries); the only method sensitive to Mars-mass
planets in the habitable zone; the only method sensitive to old
free-floating planets; the only method that is sensitive to
planets independent of the mass of their hosts; and the only
method that has good sensitivity to planets in two major Galaxy
environments (disk and bulge).  That is, while microlensing 
certainly does not supersede all other methods, it is the
best single method for conducting a systematic survey of planets
as a function of planet mass, host mass, host-planet separation, 
and position in the Galaxy.

Given the strong claims just made, why has microlensing so far 
discovered only 14 planets (8 published + 6 in prep), 
while transits have discovered dozens 
and RV has discovered hundreds?  The short answer is that
to fully achieve the above potential requires a wide-field
imager in space.

Today, microlensing is already making some remarkable discoveries
about planets:  first detection of ``cold Neptunes'', 
first Sun/Jupiter/Saturn analog,
and lowest-mass planet around a ``normal'' star.
But when considering these discoveries, it is important to keep in mind that
they are being made by 1m class telescopes {\it and smaller}.
Indeed, amateurs (equipped with 25--40cm telescopes) made major
contributions to the detection of 8 out of 14 microlensing planets.

These early successes have led to funding of a second generation
of microlensing planet searches.  The first generation combines a
wide-field search for microlensing events with followup of the
most promising events by small telescopes to search for planets.
This yields 2-3 planets per year.  In the second stage,
wide-field cameras will continuously monitor about 16 $\rm deg^2$,
which will yield dozens of planets per year.  Note that while
US astronomers played a major role 
in stage one, none ({\it zero}) of the roughly
\$40M required to build the second generation is coming from US
sources.

The third generation is a space-based wide-field imager.  Detailed
simulations show that it will improve both the mass limit and the
number of detections by an order of magnitude relative to stage two.  
Again, it is not
my purpose here to review these simulations or designs, but to
try to push the thinking of the committee ``out of the box''.

\section{Bottom Line
\label{conclusion}}

DE science will be best served by a mission that can be launched
quickly and so can obtain early results, stimulate new theoretical
work, as well as new scientific and engineering ideas on how
to proceed to the next step.

No convincing argument can be made for attacking DE first by
WL, BAO, or SN.  All have merits and demerits. 

However, a WL mission will simultaneously enable a search for
microlensing planets that will revolutionize the field
of extra-solar planets.

The 2010 Decadal Committee cannot tread in the path
charted by 2000 Committee of trying to be ``all things to all men''. 
As bad as that path proved to be in the boom years of the present decade,
it would lead to complete catastrophe in the next one, which will be
subject to much more severe financial constraints.

In DE (and probably most other areas as well), the Committee must
chart a course of faster, cheaper and ``better''.

\end{document}